\documentclass[a4paper,11pt]{article}
\usepackage{pos}
\usepackage{booktabs}
\usepackage{float}
\usepackage{subcaption}

\title{Hadronic contributions to $\alpha(Q^{2})$ and $\sin^{2}\theta_\text{W}(Q^{2})$ from spectral reconstruction of lattice-QCD data}
\ShortTitle{Hadronic contributions to $\alpha(Q^2)$ and $\sin^2\theta_\text{W}(Q^2)$}

\author*[1,2]{Adrián del Pino}
\author[3]{David A. Clarke}
\author[3]{Carleton DeTar}
\author[4,5]{Aida X.\ El-Khadra}
\author[6]{Elvira Gámiz}
\author[7]{Steven Gottlieb}
\author[8]{Anthony V. Grebe}
\author[7]{Leon Hostetler}
\author[9]{William I. Jay}
\author[8]{Andreas S. Kronfeld}
\author[3]{Shaun Lahert}
\author[10]{Jack Laiho}
\author[4,5]{Michael Lynch}
\author[4,5]{Andrew T. Lytle}
\author[11]{Ethan T. Neil}
\author[12]{Curtis T. Peterson}
\author[8]{James N. Simone}
\author[12]{Jacob W. Sitison}
\author[8]{Ruth S. Van de Water}
\author[1,2]{Alejandro Vaquero}

\affiliation[1]{Departamento de Física Teórica, Universidad de Zaragoza, 50009 Zaragoza, Spain}
\affiliation[2]{Center for Astroparticles and High Energy Physics (CAPA), Calle Pedro Cerbuna 12, 50009 Zaragoza, Spain}
\affiliation[3]{Department of Physics and Astronomy, University of Utah, Salt Lake City, UT 84112, USA}
\affiliation[4]{Department of Physics, University of Illinois, Urbana, Illinois, 61801, USA}
\affiliation[5]{Illinois Center for Advanced Studies of the Universe, University of Illinois, Urbana, Illinois, 61801, USA}
\affiliation[6]{CAFPE and Departamento de Física Teórica y del Cosmos, Universidad de Granada, E-18071 Granada, Spain}
\affiliation[7]{Department of Physics, Indiana University, Bloomington, Indiana 47405, USA}
\affiliation[8]{Theory Division, Fermi National Accelerator Laboratory, Batavia, Illinois, 60510, USA}
\affiliation[9]{Department of Physics, Colorado State University, Fort Collins, CO 80523, USA}
\affiliation[10]{Department of Physics, Syracuse University, NY 13244, USA}
\affiliation[11]{Department of Physics, University of Colorado, Boulder, Colorado 80309, USA}
\affiliation[12]{Department of Computational Mathematics, Science and Engineering, and Department of Physics and Astronomy, Michigan State University, East Lansing, Michigan 48824, USA}

\affiliation{\textbf{\textsf{Fermilab Lattice and MILC Collaborations}}}

\abstract{
    We present preliminary results from a lattice-QCD study of the hadronic contributions to the running of the electromagnetic coupling, $\Delta\alpha(Q^2)$, and the electroweak mixing angle, $\Delta\sin^2\theta_\text{W}(Q^2)$. Using $N_f = 2+1+1$ HISQ ensembles at physical quark masses, we discuss the challenges posed by strong statistical correlations in the time-momentum representation and propose a spectral-reconstruction strategy to obtain controlled continuum-extrapolated results across the full energy range.
}

\FullConference{The 42nd International Symposium on Lattice Field Theory (LATTICE 2025)\\
27 July - 1 August 2025\\
Mumbai, India\\}

\begin{document}
\maketitle

\tableofcontents
\newpage

\section{Introduction and motivation}

The precision of Standard Model electroweak tests relies heavily on the accurate determination of the running of fundamental couplings, specifically the electromagnetic coupling $\alpha(Q^2)$ and the electroweak mixing angle $\sin^2\theta_\text{W}(Q^2)$~\cite{PDG2023, WhitePaper2020}. The running of these parameters is driven by the vacuum polarization function, where the hadronic contribution represents a dominant source of uncertainty~\cite{Conigli_2025, Risch_2024, Bazavov_2025, Bazavov_PRL2025}.

Traditionally, these contributions are estimated through a phenomenological approach using experimental input from $e^+e^- \rightarrow \text{hadrons}$ and dispersion relations~\cite{Jegerlehner2023, Davier_2020}. However, the lattice QCD approach offers a first-principles alternative that allows access to individual flavor contributions and provides a completely independent alternative~\cite{FLAG_2021}. 

In this contribution, we report on a lattice-QCD calculation using the HISQ ensembles at physical quark masses to determine $\Delta\alpha(Q^2)$ and $\Delta\sin^2\theta_\text{W}(Q^2)$. While the time-momentum representation (TMR)~\cite{TMR_Bernecker} is a convenient framework, it presents challenges regarding statistical correlations across different $Q^2$ values, which complicate a simultaneous continuum extrapolation of the entire running curve. To overcome these limitations and obtain controlled, continuous results across the full energy range, we introduce an alternative strategy based on the reconstruction of smeared spectral densities~\cite{HLT_2019}. Beyond resolving these numerical instabilities, this continuous determination provides a vital strategic benchmark for the upcoming MUonE experiment~\cite{Abbiendi_2024,MUonE_2024}, which aims to measure $\Delta\alpha$ at space-like momenta across a broad range of energy scales. This dual-method approach allows for a comprehensive validation against future experimental data, offering a key input for testing the consistency of the Standard Model at the high-precision frontier.

\section{Theoretical framework}

The scale dependence of the fundamental couplings in the Standard Model is a direct consequence of vacuum polarization effects. In the case of the electromagnetic coupling, the relationship between the fine-structure constant at zero momentum, $\alpha$, and its effective value at a energy scale $Q^2$ is typically expressed as:
\begin{equation}
    \alpha(Q^2) = \frac{\alpha(0)}{1 - \Delta\alpha(Q^2)},
\end{equation}
where the shift $\Delta\alpha(Q^2)$ encapsulates the contributions from leptonic and hadronic loops. While leptonic contributions are known to high precision via perturbation theory, the hadronic part, $\Delta\alpha_\text{had}(Q^2)$, involves low-energy QCD dynamics that require non-perturbative methods like lattice QCD. 

A similar logic applies to the electroweak mixing angle, $\sin^2\theta_\text{W}$, which parametrizes the mixing between the $SU(2)_L$ and $U(1)_Y$ gauge sectors. Its running is defined as~\cite{Erler2018}
\begin{equation}
    \sin^2\theta_\text{W}(Q^2) = \sin^2\theta_\text{W} \left[ 1 + \Delta\sin^2\theta_\text{W}(Q^2) \right].
\end{equation}

\subsection{Vacuum polarization and hadronic contributions}

The leading hadronic contributions to these running observables are proportional to the subtracted vacuum polarization functions $\hat{\Pi}^{XX}(Q^2)$, where the index $XX$ denotes the specific current-current channel, namely $\gamma\gamma$ for the electromagnetic coupling and $Z\gamma$ for the electroweak mixing angle. The relations are:
\begin{align}
    \Delta\alpha_\text{had}(Q^{2}) &= 4\pi \alpha \hat{\Pi}^{\gamma\gamma}(Q^{2}), \\
    [\Delta \sin^{2}\theta_\text{W}(Q^{2})]_\text{had} &= -\frac{4\pi \alpha}{\sin^{2}\theta_\text{W}} \hat{\Pi}^{Z\gamma}(Q^{2}),
\end{align}
where $\hat{\Pi}^{XX}(Q^{2}) = \Pi^{XX}(Q^{2}) - \Pi^{XX}(0)$ ensures that the contribution vanishes at $Q^2 = 0$.\footnote{We work in the Euclidean space, where $Q^2 > 0$ represents the squared momentum transfer.}

In the lattice regularization, these functions are computed using the TMR~\cite{TMR_Bernecker}. 
This framework allows us to relate the polarization function to the zero-momentum vector-vector correlator $G^{XX}(t)$ through a weighted integral:
\begin{equation}
    \hat{\Pi}^{XX}(Q^{2}) = \int_{0}^{\infty} dt \, G^{XX}(t) K(t, Q^2),
\end{equation}
where the TMR kernel is given by $K(t, Q^2) = t^{2} - (4/Q^{2}) \sin^{2}(Qt/2)$. 
The correlator $G^{XX}(t)$ is obtained from the spatial components of the corresponding currents $j_k^X$~\cite{Francis_2014, DellaMorte_2017}:
\begin{equation}
    G^{XX}(t) = -\frac{1}{3} \int d^{3}x \sum_{k=1}^{3} \langle j_{k}^{X}(\vec{x}, t) j_{k}^{X}(0) \rangle.
\end{equation}

\subsection{Flavor decomposition and currents}

To achieve high precision, we decompose the correlators into their individual flavor components. For the electromagnetic (photon) channel, the total correlator is a sum of light ($l$), strange ($s$), and charm ($c$) contributions, weighted by their respective electric charges:
\begin{equation}
    G^{\gamma\gamma}(t) = \frac{5}{9}C^{ll}(t) + \frac{1}{9}C^{ss}(t) + \frac{4}{9}C^{cc}(t) - \frac{1}{9}D^{ls}(t),
\end{equation}
where $C^{ff}$ denotes the connected contributions and $D^{ls}$ represents the flavor-singlet disconnected terms. 

The $Z\gamma$ channel, which is relevant for the running of the electroweak mixing angle, presents a different weighting scheme due to the coupling of the $Z$ boson to the axial and vector hadronic currents. The resulting correlator $G^{Z\gamma}(t)$ is expressed as:
\begin{equation}
    \begin{split}
    G^{Z\gamma}(t) &= \left(\frac{3}{12} - \frac{5}{9}\sin^2\theta_\text{W}\right)C^{ll}(t) + \left(\frac{1}{12} - \frac{1}{9}\sin^2\theta_\text{W}\right)C^{ss}(t) + \left(\frac{2}{12} - \frac{4}{9}\sin^2\theta_\text{W}\right)C^{cc}(t)\\
    &+ \frac{1}{9}\sin^2\theta_\text{W} D^{ls}(t).
\end{split}
\end{equation}

These detailed decompositions allow us to isolate the systematic effects associated with each flavor, such as the different signal-to-noise ratios and the impact of the disconnected contributions, which, although suppressed, are crucial for a complete first-principles determination. 

It should be noted that these decompositions assume isospin symmetry. Isospin-breaking effects are reserved for future work. Following similar muon $g-2$ HVP analyses~\cite{Bazavov_2025,Bazavov_PRL2025}, we expect these corrections to be at the $\sim 0.01\%$ level.

\section{Ensemble setup}

The numerical simulations are performed using gauge configurations generated by the MILC Collaboration. 
Our setup employs the highly improved staggered quarks (HISQ) action for both sea and valence quarks, with $N_f = 2+1+1$ dynamical flavors. 

A critical advantage of this setup is that all ensembles are tuned to physical quark masses, which significantly reduces the systematic uncertainties associated with chiral extrapolation.
To ensure a robust continuum limit, we use four different lattice spacings ranging from $a \approx 0.06$ fm to $0.15$ fm. The hadronic currents are implemented as local vector currents, renormalized by the factor $Z_V$ to match the continuum scaling. 

\begin{table} 
\centering
\caption{Ensembles used in this work.
The first column denotes the approximate lattice spacing ($a$) in fm, while the second and third columns specify the physical spatial extent ($L$) and the lattice size, respectively. The simulation sea-quark massesin lattice units are presented in the fourth through seventh columns. Finally, the eighth column provides the gradient-flow scale $w_0/a$, as defined in Ref.~\cite{FermilabMILC_Omega_2025}. To map our numerical results onto physical units, we use $w_0 = 0.1715(9)$ fm~\cite{Dowdall_2013}.}
\label{tab:ensembles}
\begin{tabular}{cccccccc}
\toprule
$\approx a/\text{fm}$ & $L/\text{fm}$ & $N_s^3 \times N_t$ & $am_u^{\text{sea}}$ & $am_d^{\text{sea}}$ & $am_s^{\text{sea}}$ & $am_c^{\text{sea}}$ & $w_0/a$ \\
\midrule
0.15   & 4.85 & $32^3 \times 48$ & 0.002426 & 0.002426 & 0.0673  & 0.8447 & 1.13227(18) \\
0.12   & 5.81 & $48^3 \times 64$ & 0.001907 & 0.001907 & 0.05252 & 0.6382 & 1.41060(28) \\
0.09 & 5.61 & $64^3 \times 96$ & 0.001326 & 0.001326 & 0.03636 & 0.4313 & 1.95021(57) \\
0.06   & 5.45 & $96^3 \times 128$ & 0.0008  & 0.0008   & 0.022   & 0.260  & 3.01838(92) \\
\bottomrule
\end{tabular}
\end{table}

\section{Analysis and preliminary results}

In this section, we present the preliminary analysis of the hadronic contributions to the vacuum polarization functions. The results shown here are preliminary and blinded. While the statistical precision is high, a full assessment of systematic uncertainties—including finite-volume effects, mass mistuning, and taste-breaking corrections—is still ongoing.

\subsection{Continuum extrapolations}

Using the four lattice spacings described in the ensemble setup ($a \approx 0.06$ to $0.15$ fm), we perform the continuum extrapolation for the individual flavor components of $\Pi(Q^2)$. To ensure a direct and robust comparison with experimental $R$-ratio determinations and other lattice results, we carry out a point-by-point continuum extrapolation for each energy scale $Q^2$ independently

To account for discretization effects and potential mistuning of the sea-quark masses, we employ a Bayesian model based on the following expansion:
\begin{equation}\hat{\Pi}(a, Q^2) = \hat{\Pi}_{\text{cont}}(Q^2) \left[ 1 + c_1 \left( \frac{a \Lambda}{\hbar c} \right)^2 + c_2 \left( \frac{a \Lambda}{\hbar c} \right)^4 + d_1 \left( \frac{a M_{D_s}}{\hbar c} \right)^2 + C_s \delta_s + \dots \right],
\end{equation}

where $a$ is the lattice spacing and the scale $\Lambda$ is chosen to maximize the Gaussian Bayes factor \cite{Lepage_2002}. The terms $\delta_s$ represent the corrections for the mistuning of the sea-quark masses relative to their physical values.To ensure the robustness of the extrapolation, we systematically test the stability of the fit against higher-order artifacts. Specifically, we compare models with and without the quartic term $a^4$ ($c_2$ coefficient) and evaluate the model preference using the Bayes factor ($BF = e^{\Delta \log GBF}$). The fit coefficients are constrained using Gaussian priors of $0(1)$, with the exception of the continuum-limit parameter. For the latter, the prior is informed by the central value and the $1$-sigma uncertainty obtained from our finest ensemble. This Bayesian framework, implemented via the lsqfit package, allows us to quantify the sensitivity of the continuum limit to the inclusion of higher-order discretization terms and mass mistuning corrections.

\begin{figure} 
    \centering
    \begin{subfigure}[b]{0.49\textwidth}
        \centering
        \includegraphics[width=\textwidth]{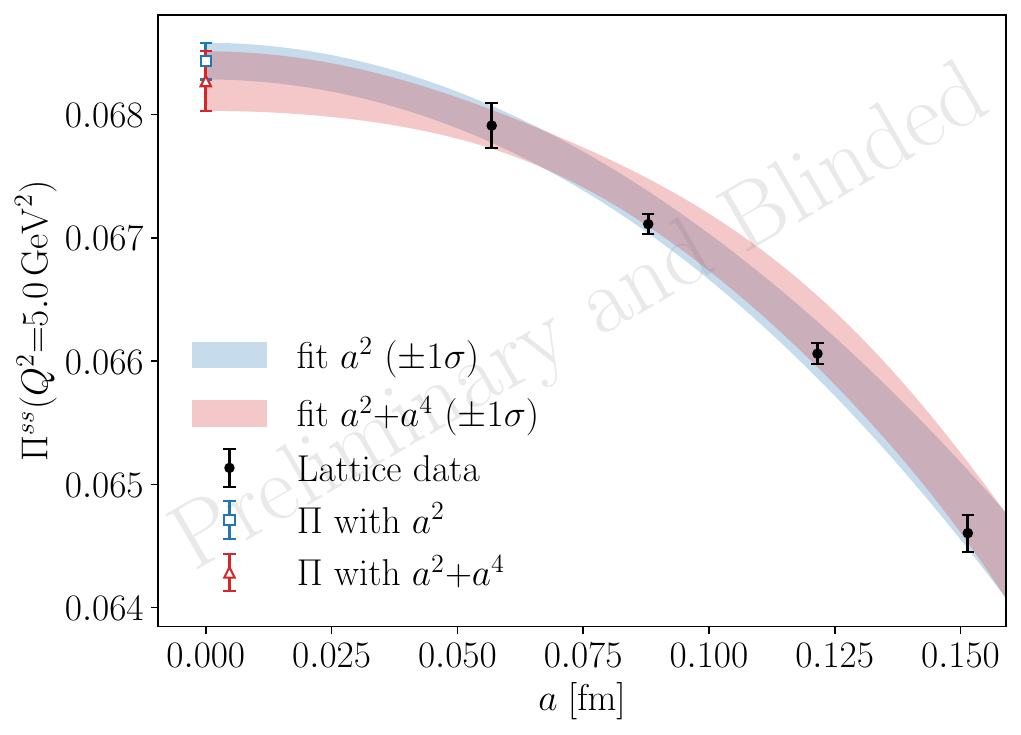}
    \end{subfigure}
    \hfill 
    \begin{subfigure}[b]{0.49\textwidth}
        \centering
        \includegraphics[width=\textwidth]{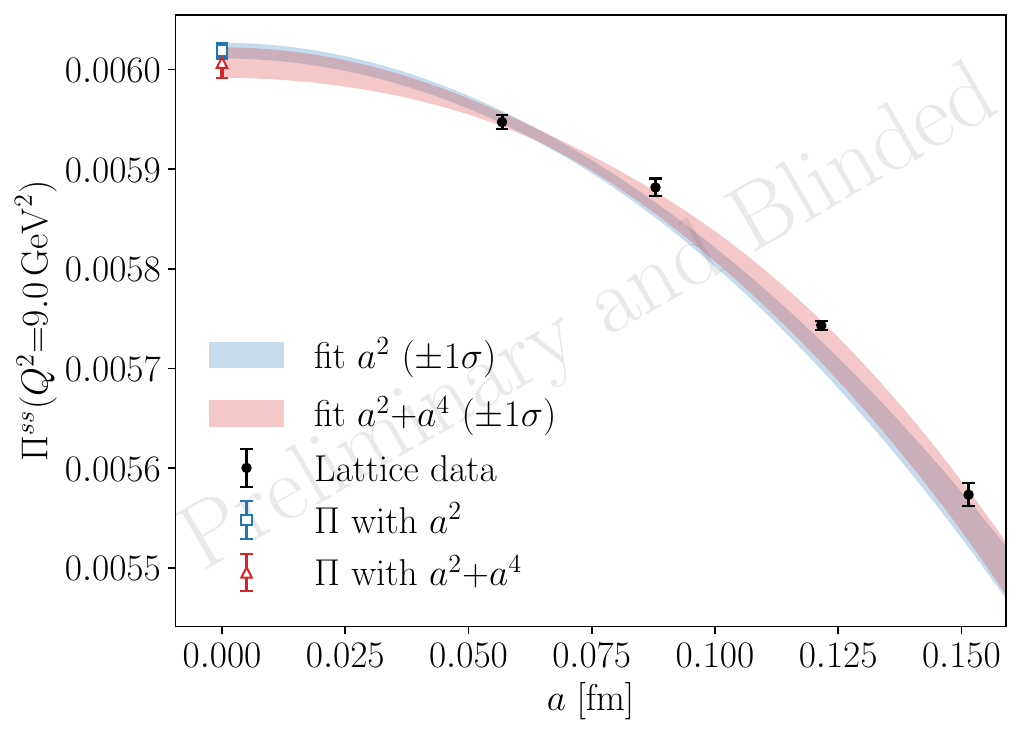}
    \end{subfigure}
    \caption{Continuum extrapolation of the light-quark contribution $\hat{\Pi}^{ll}$ at $Q^2 = 5.0 \text{ GeV}^2$ (left) and the strange-quark contribution $\hat{\Pi}^{ss}$ at $Q^2 = 9.0 \text{ GeV}^2$ (right). Data points are compared against different fit forms including $a^2$ and $a^4$ terms. The error bars represent statistical uncertainties only. All results are preliminary and blinded.}
    \label{fig:extrapolation}
\end{figure}

In Figure~\ref{fig:extrapolation}, we illustrate the extrapolation for the light-quark ($\Pi^{ll}$) and strange-quark ($\Pi^{ss}$) contributions at specific $Q^2$ values. We test different fit forms, specifically including $a^2$ and $a^4$ terms to ensure stability against lattice artifacts.

\subsection{Running of \texorpdfstring{\boldmath$\alpha(Q^2)$}{alpha} and \texorpdfstring{\boldmath$\sin^2\theta_\text{W}(Q^2)$}{sin2thetaW}}

\begin{figure} 
    \centering
    \begin{subfigure}[b]{0.49\textwidth}
        \centering
        \includegraphics[width=\textwidth]{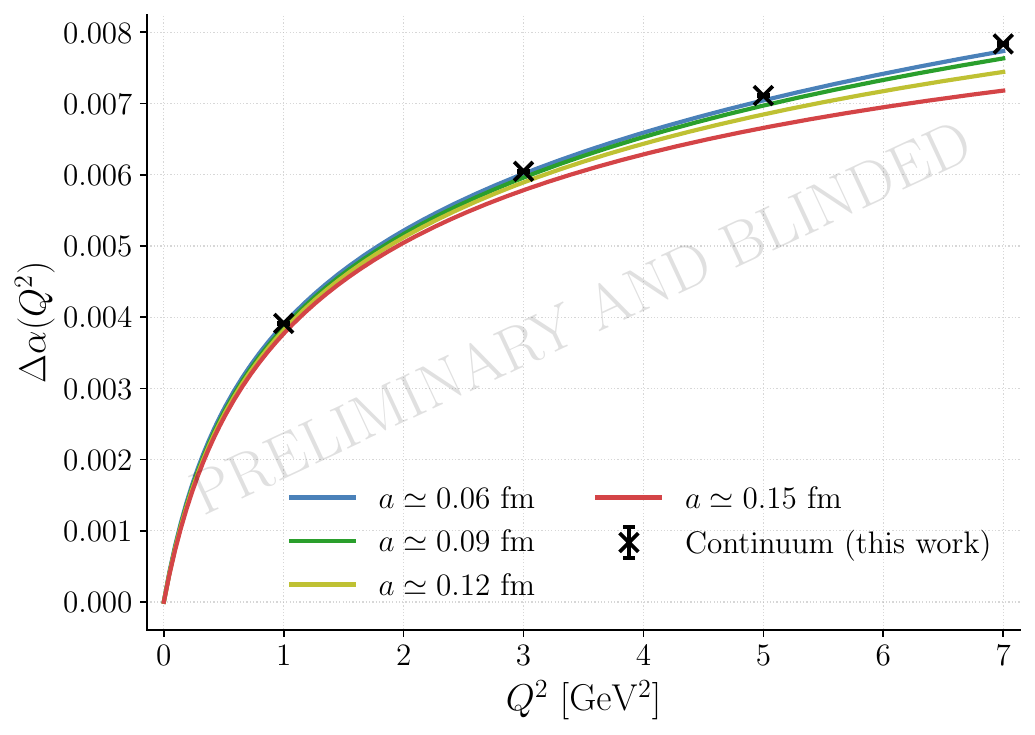}
    \end{subfigure}
    \hfill 
    \begin{subfigure}[b]{0.49\textwidth}
        \centering
        \includegraphics[width=\textwidth]{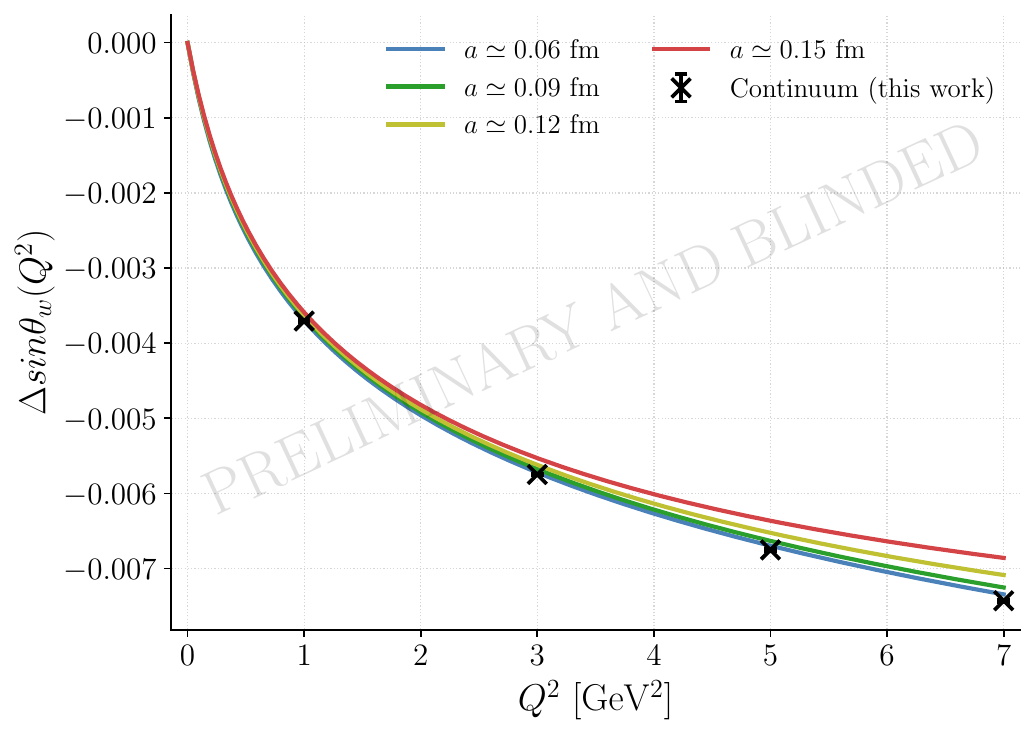}
    \end{subfigure}
    \caption{Running of the electromagnetic coupling $\Delta\alpha(Q^2)$ (left) and the electroweak mixing angle $\Delta\sin^2\theta_\text{W}(Q^2)$ (right) as a function of the energy scale $Q^2$. The colored lines represent the results from our lattice ensembles at various spacings, while the black circles correspond to the point-by-point continuum extrapolated values. All error bars represent statistical uncertainties only. All results are preliminary and currently blinded.}
    \label{fig:results}
\end{figure}

The flavor-decomposed vacuum polarization functions $\hat{\Pi}^{XX}(Q^2)$, obtained from the point-by-point continuum extrapolation described in the previous section, form the basis for our final determinations. By applying the specific weighting schemes for the electromagnetic and electroweak mixing channels, we compute the running of the electromagnetic coupling, $\Delta\alpha(Q^2)$, and the electroweak mixing angle, $\Delta\sin^2\theta_\text{W}(Q^2)$, as presented in Figure~\ref{fig:results}.

Our blinded analysis achieves high statistical precision for $0 < Q^2 < 7 \text{ GeV}^2$. This continuum-extrapolated data aims to provide a definitive, first-principles contribution to global evaluations of these couplings. Presently, errors are purely statistical; systematic uncertainties are being evaluated for the final error budget.

\section{Spectral-density reconstruction}

\begin{figure} 
    \centering
    \begin{subfigure}[b]{0.49\textwidth}
        \centering
        \includegraphics[width=\textwidth]{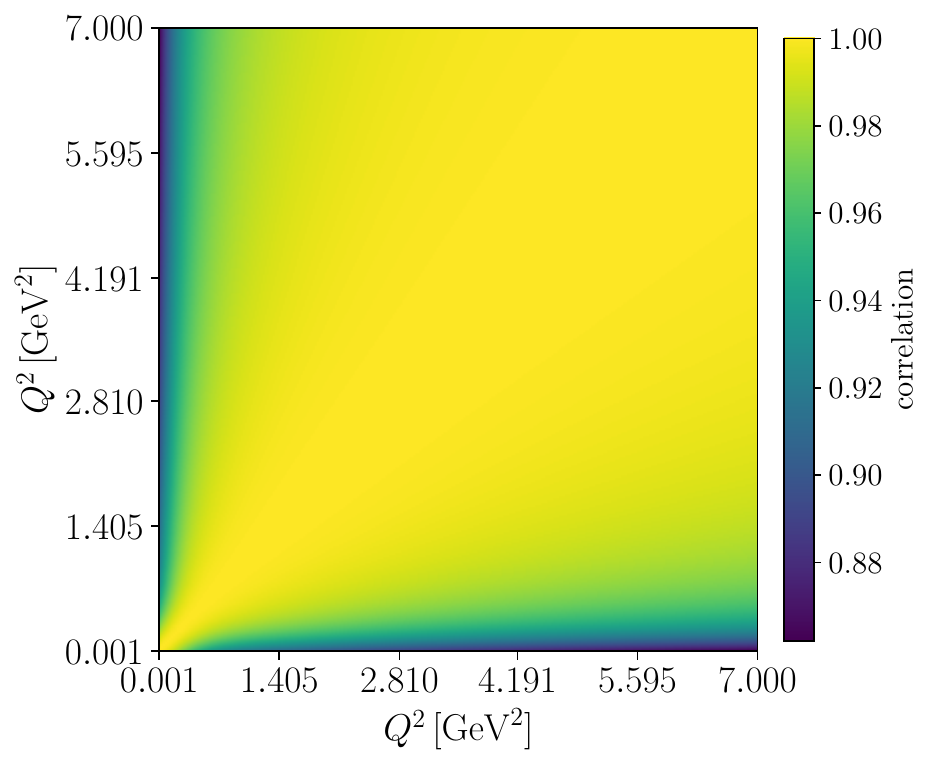}
    \end{subfigure}
    \hfill 
    \begin{subfigure}[b]{0.49\textwidth}
        \centering
        \includegraphics[width=\textwidth]{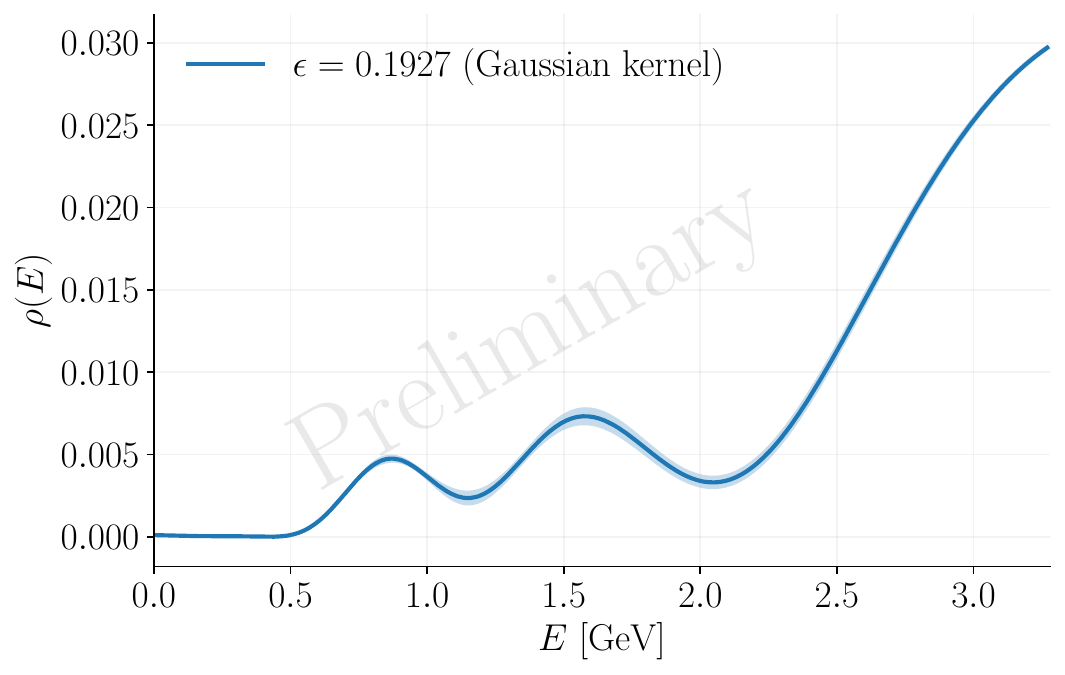}
    \end{subfigure}
    \caption{(Left) Correlation matrix of $\hat{\Pi}(Q^2)$ on the $a \approx 0.06$ fm ensemble, highlighting the strong off-diagonal correlations that indicate a nearly singular matrix. (Right) Preliminary reconstruction of the smeared spectral density $\rho_{\epsilon}(E)$ using the HLT method with a smearing width of $\epsilon = 0.1927$ GeV on the same ensemble. Error bars represent statistical uncertainties only.}
    \label{fig:spectral}   
\end{figure}

Although point-by-point extrapolations reliably yield $\Delta\alpha(Q^2)$ and $\Delta\sin^2\theta_\text{W}(Q^2)$ at discrete scales, we seek continuous curves. Technically, the TMR strongly correlates different $Q^2$ points due to shared low-energy contributions from the vector correlator $G(t)$. This produces a nearly singular correlation matrix (Fig.~\ref{fig:spectral}), rendering simultaneous continuum extrapolations of the full curve unstable.

Beyond mitigating numerical instabilities, a continuous $\Delta\alpha(Q^2)$ directly benchmarks the upcoming MUonE experiment~\cite{Abbiendi_2024, Lautrup_2015}, which measures space-like $\Delta\alpha$ over broad energy scales. Unlike discrete evaluations, our approach enables comprehensive validation against MUonE data, providing crucial input for high-precision Standard Model tests.

To bridge these requirements and overcome the TMR induced correlations, we adopt an alternative strategy: the reconstruction of smeared spectral densities $\rho_{\epsilon}(E)$. The relationship between the zero-momentum correlator $G(t)$ and the spectral density is given by the Laplace transform:
\begin{equation}
G(t) = \int_{0}^{\infty} dE  \rho(E) e^{-Et},
\end{equation}
allowing the polarization function to be expressed as:
\begin{equation}
\hat{\Pi}(Q^{2}) = \int_{0}^{\infty} dE  \rho(E) W(E, Q^2), \quad\quad W(E, Q^2) = \frac{2Q^2}{E^3(E^2 + Q^2)}.
\end{equation}

We use the Hansen-Lupo-Tantalo (HLT) method~\cite{HLT_2019} to extract smeared spectral densities, $\rho_{\epsilon}(E)$. This reconstructs the spectral density by balancing the approximation of the energy kernel $W(E, Q^2)$ (the Laplace transform of the TMR kernel $K(t, Q^2)$) against lattice statistical noise.

The HLT method operates by constructing a smeared spectral density, $\rho_{\epsilon}(E)$, which is defined as a convolution of the true spectral function with a resolution function (or kernel) of width $\epsilon$:
\begin{equation}
\rho_{\epsilon}(E) = \int_{0}^{\infty} dE' \rho(E') \Delta_{\epsilon}(E, E'),
\end{equation}
where $\Delta_{\epsilon}$ is typically chosen to be a Gaussian or similar peaking function. The core of the HLT approach is to approximate the target weight function $W(E, Q^2)$, which is required to compute $\hat{\Pi}(Q^2)$, using a linear combination of basis functions $\delta_t(E) = e^{-Et}$ that correspond to the correlator data $G(t)$:
\begin{equation}
W_{\Delta}(E) = \sum_{t} g_t(Q^2) e^{-Et} \approx W(E, Q^2).
\end{equation}
The coefficients $\{g_t\}$ are determined by minimizing a functional that incorporates both the approximation error (the difference between $W_{\Delta}$ and $W$) and a regularization term proportional to the statistical covariance of the correlator.

Adjusting the width $\epsilon$ controls the bias-variance tradeoff: a smaller $\epsilon$ improves the target kernel approximation but amplifies statistical noise. We first extrapolate the smeared densities to the continuum at fixed $\epsilon$, and then take the unsmeared limit ($\epsilon \to 0$) to recover the physical spectral function. This two-step approach keeps systematic reconstruction uncertainties strictly controlled.

A preliminary reconstruction is shown in Fig.~\ref{fig:spectral}; future work will involve a controlled extrapolation in the smearing width $\epsilon$ to reach the unsmeared limit.

\section{Summary and outlook}

In this contribution, we have presented preliminary and blinded results from a lattice QCD study of the hadronic contributions to the running of the electromagnetic coupling, $\Delta\alpha(Q^{2})$, and the electroweak mixing angle, $\Delta\sin^{2}\theta_\text{W}(Q^{2})$. Using $N_f = 2+1+1$ HISQ ensembles at physical quark masses, we have demonstrated that the data provide high statistical precision across multiple lattice spacings.

An analysis based solely on TMR has a significant challenge in the standard time-momentum representation approach: the strong statistical correlations between different $Q^2$ points. These correlations, driven by the low-energy tail of the vector correlators, make a simultaneous continuum extrapolation for the entire running curve difficult. To address this, we are developing a robust strategy based on spectral reconstruction using the Hansen-Lupo-Tantalo (HLT) method. 
This dual-method approach not only allows for a continuous determination of $\hat{\Pi}(Q^2)$ with better-controlled systematic uncertainties but also provides a vital strategic benchmark for the upcoming MUonE experiment, facilitating a direct comparison with experimental data at space-like momenta.

Moving forward, our immediate goals include:
\begin{itemize}
    \item The inclusion of disconnected contributions ($D^{ls}$) to complete the flavor decomposition 
    \item A comprehensive assessment of systematic corrections, including finite-volume effects, mass mistuning, and taste-breaking artifacts.
    \item A detailed investigation of the low-energy region ($Q^2 < 1 \text{ GeV}^2$), where the sensitivity of the kernels is most pronounced.
    \item The final determination of $\Delta\alpha(M_Z^2)$ and $\Delta\sin^2\theta_\text{W}(M_Z^2)$ to provide a first-principles cross-check against experimental and phenomenological results.
\end{itemize}
Ultimately, we aim to deliver a precise and competitive comparison between the TMR and spectral reconstruction methods, contributing to the global effort of testing the Standard Model.

\section*{Acknowledgments}

Computations for this work were carried out in part with computing and long-term storage resources provided by the USQCD Collaboration, the National Energy Research Scientific Computing Center (Cori), the Argonne Leadership Computing Facility (Mira) under the INCITE program, and the Oak Ridge Leadership Computing Facility (Summit) under the Innovative and Novel Computational Impact on Theory and Experiment (INCITE) and the ASCR Leadership Computing Challenge (ALCC) programs, which are funded by the Office of Science of the U.S. Department of Energy. This work used the Extreme Science and Engineering Discovery Environment (XSEDE) supercomputer Stampede 2 at the Texas Advanced Computing Center (TACC) through allocation TG-MCA93S002. The XSEDE program is supported by the National Science Foundation under grant number ACI-1548562. Computations on the Big Red II+, Big Red 3, and Big Red 200 supercomputers were supported in part by Lilly Endowment, Inc., through its support for the Indiana University Pervasive Technology Institute. The parallel file system employed by Big Red II+ was supported by the National Science Foundation under Grant No.\ CNS-0521433. This work utilized the RMACC Summit supercomputer, which is supported by the National Science Foundation (awards ACI-1532235 and ACI-1532236), the University of Colorado Boulder, and Colorado State University. The Summit supercomputer is a joint effort of the University of Colorado Boulder and Colorado State University. Some of the computations were done using the Blue Waters sustained-petascale computer, which was supported by the National Science Foundation (awards OCI-0725070 and ACI-1238993) and the state of Illinois. Blue Waters was a joint effort of the University of Illinois at Urbana-Champaign and its National Center for Supercomputing Applications. We also used the Cambridge Service for Data Driven Discovery (CSD3), part of which is operated by the University of Cambridge Research Computing Service on behalf of the Science and Technology Facilities Council (STFC) DiRAC HPC Facility. The DiRAC component of CSD3 was funded by BEIS capital funding via STFC capital grants ST/P002307/1 and ST/R002452/1 and STFC operations grant ST/R00689X/1.

This work was supported in part by the U.S. Department of Energy, Office of Science, under Awards No.\ DE-SC0010005 (E.T.N. and J.W.S.), No.\ DE-SC0010120 (S.G.), No.\ DE-SC0015655 (A.X.K., S.L., M.L., A.T.L.), No.\ DE-SC0009998 (J.L.), the ``High Energy Physics Computing Traineeship for Lattice Gauge Theory'' No.\ DE-SC0024053 (J.W.S.), and the Funding Opportunity Announcement Scientific Discovery through Advanced Computing: High Energy Physics, LAB 22-2580 (D.A.C., C.T.P., L.H., M.L., S.L.); by the Exascale Computing Project (17-SC-20-SC), a collaborative effort of the U.S. Department of Energy Office of Science and the National Nuclear Security Administration (H.J.); by the National Science Foundation under Grants Nos. PHY20-13064 and PHY23-10571 (C.D., D.A.C., S.L., A.V.), PHY23-09946 (A.B.), Grant No.\ 2139536 for Characteristic Science Applications for the Leadership Class Computing Facility (L.H., H.J.), and Grant No.\ PHY-2309135 to the Kavli Institute for Theoretical Physics (KITP); by MICIU/AEI/10.13039/501100011033 and FEDER (EU) under Grant PID2022-140440NB-C21 (E.G.); by Consejería de Universidad, Investigación e Innovación and Gobierno de España and EU--NextGenerationEU, under Grant AST22 8.4 (E.G.); by AEI (Spain) under Grant No.\ RYC2020-030244-I / AEI / 10.13039/501100011033 (A.V.). A.X.K. and E.T.N. are grateful to the Pauli Center for Theoretical Studies and the ETH Zürich for support and hospitality.  
This document was prepared by the Fermilab Lattice and MILC Collaborations using the resources of the Fermi National Accelerator Laboratory (Fermilab), a U.S.\ Department of Energy, Office of Science, HEP User Facility.
Fermilab is managed by Fermi Forward Discovery Group, LLC, acting under Contract No.~89243024CSC000002 with the U.S.\ Department of Energy.


\begin{thebibliography}{99}

\bibitem{PDG2023}
R.L. Workman et al. (Particle Data Group), 
\textit{Review of particle physics}, 
\textit{Prog. Theor. Exp. Phys.} \textbf{2022}, 083C01 (2022) and 2023 update.

\bibitem{WhitePaper2020}
T. Aoyama et al., 
\textit{The theory of the muon $g-2$: status and precision priorities}, 
\textit{Phys. Rept.} \textbf{887}, 1-166 (2020) [arXiv:2006.04822].

\bibitem{Conigli_2025}
A. Conigli et al.,
\textit{Precision lattice calculation of the hadronic contribution to the running of the electroweak gauge couplings},
arXiv:2511.01623 [hep-lat] (2025).

\bibitem{Risch_2024}
A. Risch et al.,
\textit{Lattice QCD calculation of the hadronic vacuum polarization contribution to the muon g-2},
arXiv:2401.04049 [hep-lat] (2024).

\bibitem{Bazavov_2025}
A. Bazavov et al. (Fermilab Lattice, HPQCD, and MILC Collaborations), 
\textit{Hadronic vacuum polarization for the muon $g-2$ from lattice QCD: Complete short and intermediate windows}, 
\textit{Phys. Rev. D} \textbf{111}, 094508 (2025) [arXiv:2411.09656].

\bibitem{Bazavov_PRL2025}
A. Bazavov et al. (Fermilab Lattice, HPQCD, and MILC Collaborations), 
\textit{Hadronic vacuum polarization for the muon $g-2$ from lattice QCD: Long-distance and full light-quark connected contribution}, 
\textit{Phys. Rev. Lett.} \textbf{135}, 011801 (2025) [arXiv:2412.18491].

\bibitem{Jegerlehner2023}
F. Jegerlehner, 
\textit{The running fine structure constant $\alpha(E)$ and the role of the hadronic contribution}, 
\textit{Found. Phys.} \textbf{53}, 35 (2023) [arXiv:2212.01014].

\bibitem{Davier_2020}
M. Davier, A. Hoecker, B. Malaescu and Z. Zhang, 
\textit{A new evaluation of the hadronic vacuum polarisation contributions to the muon $g-2$ and to $\alpha(m_Z^2)$}, 
\textit{Eur. Phys. J. C} \textbf{80}, 241 (2020) [arXiv:1908.00921].

\bibitem{FLAG_2021}
Y. Aoki et al. (FLAG Collaboration), 
\textit{FLAG review 2021: Results in lattice QCD configured for particle physics}, 
\textit{Eur. Phys. J. C} \textbf{82}, 869 (2022) [arXiv:2111.09849].

\bibitem{TMR_Bernecker}
H. Bernecker and H. B. Meyer, 
\textit{Vector correlators in lattice QCD: Methods and applications}, 
\textit{Eur. Phys. J. A} \textbf{47}, 148 (2011) [arXiv:1107.4388].

\bibitem{HLT_2019}
M. Hansen, A. Lupo, and N. Tantalo, 
\textit{Extraction of spectral densities from lattice correlators}, 
\textit{Phys. Rev. D} \textbf{99}, 094508 (2019) [arXiv:1903.06476].

\bibitem{MUonE_2024}
A. Abbiendi et al. (MUonE Collaboration), 
\textit{Proposal for Phase~1 of the MUonE experiment}, 
CERN-SPSC-2024-015 / SPSC-P-370 (April 2024).

\bibitem{Abbiendi_2024}
G. P. Abbiendi et al., 
\textit{An alternative evaluation of the leading-order hadronic contribution to the muon $g-2$ with MUonE}, 
\textit{Phys. Lett. B} \textbf{848}, 138344 (2024) [arXiv:2309.14205].

\bibitem{Erler2018}
J. Erler and R. Ferro-Hernández, 
\textit{Weak mixing angle in the Thomson limit}, 
\textit{JHEP} \textbf{03}, 196 (2018) [arXiv:1712.08704].

\bibitem{Francis_2014}
A. Francis, B. Jaeger, H.B. Meyer and H. Wittig, 
\textit{A new representation of the Adler function for lattice QCD}, 
\textit{Phys. Rev. D} \textbf{90}, 011501 (2014) [arXiv:1403.1378].

\bibitem{DellaMorte_2017}
M. Della Morte et al., 
\textit{The hadronic vacuum polarization contribution to the muon $g-2$ from lattice QCD}, 
\textit{JHEP} \textbf{10}, 020 (2017) [arXiv:1705.01775].



\bibitem{FermilabMILC_Omega_2025}
A. Bazavov et al. (Fermilab Lattice and MILC Collaborations), 
\textit{High-Precision Scale Setting with the Omega-Baryon Mass and Gradient Flow}, 
(2025) [arXiv:2509.14367].

\bibitem{Dowdall_2013}
R. J. Dowdall, C. T. H. Davies, G. P. Lepage, and C. McNeile (HPQCD Collaboration), 
\textit{$V-A$ correlators in lattice QCD with physical-mass HISQ quarks}, 
\textit{Phys. Rev. D} \textbf{88}, 074504 (2013) [arXiv:1303.1670].

\bibitem{Lepage_2002}
G. P. Lepage et al., 
\textit{Constrained curve fitting in lattice QCD}, 
\textit{Nucl. Phys. B Proc. Suppl.} \textbf{106}, 12 (2002) [arXiv:hep-lat/0110175].

\bibitem{Lautrup_2015}
B. Lautrup et al.,
\textit{A new approach to evaluate the leading hadronic corrections to the muon $g-2$},
\textit{Phys. Lett. B} \textbf{746}, 325 (2015) [arXiv:1504.02228].


\end{thebibliography}
\end{document}